\documentclass[a4paper,12pt]{article}
\pdfoutput=1
\usepackage{amsmath,amsfonts,amssymb,epsfig,comment}

\numberwithin{equation}{section}

\usepackage{youngtab}

\DeclareSymbolFont{extraup}{U}{zavm}{m}{n}
\DeclareMathSymbol{\vardiamond}{\mathalpha}{extraup}{87}

\newcommand{\thetab}{\bar{\theta}}

\def\twomat[#1,#2][#3,#4]{\left( \begin{array}{cc} #1 & #2 \\ #3 & #4 \end{array} \right)}
\def\threemat[#1,#2,#3][#4,#5,#6][#7,#8,#9]{\left( \begin{array}{ccc} #1 & #2 & #3\\ #4 & #5 & #6 \\ #7 & #8 & #9 \end{array} \right)}
\def\twovec[#1,#2]{\left( \begin{array}{c} #1  \\ #2 \end{array} \right)}

\def\lagr{\mathcal{L}}

\newcommand{\bea}{\begin{eqnarray}}
\newcommand{\eea}{\end{eqnarray}}

\newcommand{\exclude}[1]{}

\def\pd[#1]{\frac{\partial}{\partial #1}}
\def\pdd[#1,#2]{\frac{\partial #1}{\partial #2}}

\def\nn{\nonumber}

\def\beq{\begin{equation}}
\def\eeq{\end{equation}}

\def\dalpha{{\dot{\alpha}}}

\def\2b2[#1,#2][#3,#4]{\left( \begin{array}{cc} #1 & #2 \\ #3 & #4 \end{array} \right)}
\def\3b3[#1,#2,#3][#4,#5,#6][#7,#8,#9]{\left( \begin{array}{ccc} #1 & #2 #3 \\ #4 & #5 & #6\\#7&#8&#9\end{array} \right)}
\def\vec2[#1,#2]{\left( \begin{array}{c} #1 \\ #2 \end{array} \right)}

\def\ov{\overline}

\def\psqr#1#2{{\vcenter{\vbox{\hrule height.#2pt
        \hbox{\vrule width.#2pt height#1pt \kern#1pt
        \vrule width.#2pt}
        \hrule height.#2pt \hrule height.#2pt
        \hbox{\vrule width.#2pt height#1pt \kern#1pt
        \vrule width.#2pt}
        \hrule height.#2pt}}}}
\def\sqr#1#2{{\vcenter{\vbox{\hrule height.#2pt
        \hbox{\vrule width.#2pt height#1pt \kern#1pt
        \vrule width.#2pt}
        \hrule height.#2pt}}}}

\title{XXX ...}

\begin{document}

\begin{flushright}
\end{flushright}
\begin{center}

\vspace{1cm}
{\LARGE{\bf Minimal constrained superfields and the Fayet-Iliopoulos model
 }}

\vspace{1cm}

\large{  Karim Benakli$^\spadesuit$ \let\thefootnote\relax\footnote{$^\spadesuit$kbenakli@lpthe.jussieu.fr},
Yifan Chen$^\clubsuit$ \footnote{$^\clubsuit$yifan.chen@lpthe.jussieu.fr},
Mark~D.~Goodsell$^\vardiamond$ \footnote{$^\vardiamond$goodsell@lpthe.jussieu.fr}
 \\[5mm]}

{\small
\emph{Sorbonne Universit\'e, UPMC Univ Paris 06, UMR 7589, LPTHE, F-75005, Paris, France \\
CNRS, UMR 7589, LPTHE, F-75005, Paris, France \\
}}

\end{center}
\vspace{0.7cm}

\abstract{ We show how the necessary constraints to project out all the components of a chiral superfield except for some scalar degrees of freedom originate from simple operators in the microscopic theory.  This is in particular useful in constructing the simplest models of a goldstone boson/inflaton; or extracting the Standard Model Higgs doublet from a supersymmetric electroweak sector. We use the Fayet-Iliopoulos model as an example of the origin for the supersymmetry breaking. We consider the regime where both gauge symmetry and supersymmetry are spontaneously broken, leaving (in the decoupling limit) the goldstino as the only light mode in this sector. We show in three different ways, both in components and in superspace language, how the nilpotent goldstino superfield emerges. We then use it to write different effective operators and extract some of the consequences for the low energy spectrum.}

\newpage

\tableofcontents

\setcounter{footnote}{0}

\section{Introduction}
\label{SEC:INTRO}

Superspace and superfields are powerful tools for the construction of globally supersymmetric theories. In  \cite{Rocek:1978nb}, it was shown how they can also be used in the case where supersymmetry is only realised non-linearly \cite{Volkov:1973ix}. The goldstino is then part of a constrained superfield $X_{NL}$. In the simplest examples  \cite{Rocek:1978nb,Casalbuoni:1988xh} the latter satisfies: 
\bea
X_{NL} ^2=0
\label{squareX}
\eea
which eliminates the scalar component, the sgoldstino. In \cite{Casalbuoni:1988xh}, the constraint (\ref{squareX})  was explicitly derived by taking the sgoldstino mass to infinity. Going further  and imposing
\bea
X_{NL} \ov{D}^2\ov{X_{NL} } \propto X_{NL} 
\label{Ffixed}
\eea
fixes the scale of supersymmetry breaking, the F-term $F_X$ in $X_{NL}$ \cite{Rocek:1978nb}. This leaves then the goldstino $\tilde{\lambda}_\alpha$ as the only independent component in the superfield:
\bea
D_\alpha X_{NL} | = \sqrt{2}\tilde{\lambda}_\alpha + \cdots ; \qquad X_{NL} |  = \frac{\tilde{\lambda}_\alpha\tilde{\lambda}^\alpha}{2F_X} +\cdots
\label{goldstinoX}
\eea
In \cite{Komargodski:2009rz}, it was argued that the constraint (\ref {Ffixed}) can be too restrictive and one can instead choose to use only (\ref{squareX}) and keep as independent components of $X_{NL}$ both the goldstino superfield and the auxiliary component.  It was also conjectured that the superfield $X$ which controls
the violation of the Ferrara-Zumino supercurrent $\mathcal{J}_{\alpha \dot{\alpha}}$ conservation equation:
\begin{align}
\ov{D}^ {\dot{\alpha}} \mathcal{J}_{\alpha \dot{\alpha}} = D_\alpha X
\end{align}
flows in the infrared to the superfield $X_{NL}$, i.e. $X \rightarrow X_{NL} $.

There are other ways to embed the goldstino in a constrained superfield. The goldstino can appear as the lowest component as it was originally described in \cite{Lindstrom:1979kq,Samuel:1982uh}. One approach is to directly write the superfield in ``splitting form'' in terms of homogeneously transforming components \cite{Ivanov:1978mx,Ivanov:1982bpa,Samuel:1982uh}:
\begin{align}
\Lambda_\alpha \equiv \sqrt{2}{F_X} \theta_\alpha + \tilde{\lambda}_\alpha (z), \qquad z^\mu \equiv x^\mu - i \theta \sigma^\mu \overline{\theta} - i \frac{\sqrt{2}}{F_X}  \tilde{\lambda} (x) \sigma^\mu \bar{\theta}.
\end{align}
On the other hand, in order to make contact with the UV-origin of the fields, we can instead identify the goldstino with a spin $1/2$ component of a vector multiplet $V_{NL} $, and this is the approach we shall take here (although the two approaches can be related by a non-linear transformation of the superfields). The corresponding constraints take then the form:
\bea V_{NL} = {V_{NL} }^\dagger \label{reality} \eea
\bea V_{NL} ^2= 0 \label{nilpotency} \eea
\bea V_{NL}  \propto V_{NL}  (D^\alpha \ov{D}^2 D_\alpha + \ov{D}^\dalpha {D}^2 \ov{D}_\dalpha) V_{NL}   \label{Dfixed} \eea
and the goldstino  is obtained from the lowest component of
\bea W_{NL \alpha} = -\frac {1}{4}  \ov{D}^2 D_\alpha  V_{NL}  = \tilde{\lambda}_\alpha + \cdots   \label{Wgoldstino} \eea
These constraints are satisfied if $V_{NL}  = \ov{X_{NL} }X_{NL}/\Lambda^2 $ where the size  of the suppression scale $\Lambda$ is given by the $X_{NL}$ F-term.  Note that  $V_{NL} $ can be used either for a true D-term breaking model  or to parametrise the effects of an F-term breaking as done in \cite{Bandos:2016xyu,Cribiori:2017ngp, Buchbinder:2017qls,GarciadelMoral:2017vnz}. Here the condition (\ref{Dfixed}) appears as a consequence of (\ref{Ffixed}). An important result shown in  \cite{Lindstrom:1979kq,Samuel:1982uh},  and subsequently in \cite{Ivanov:1982bpa,Luo:2009ib,Liu:2010sk,Kuzenko:2010ef} for the other representations, is that the corresponding Lagrangian is the Volkov-Akulov one  or can be mapped to it through field redefinitions.

This nilpotent superfield construction allows to describe the coupling of the goldstino to matter fields in the lower energy effective theory. Writing appropriate constraint equations,  $X_{NL}$ allows to project out  heavy components in matter superfields without explicitly going through integrating them out in the ultaviolet (UV) theory Lagrangian \cite{Komargodski:2009rz}. We shall consider here the constraint equation:
\beq 
X_{NL} (\mathcal{A} + \overline{\mathcal{A}}) = 0, 
\label{constrintIntro}
\eeq
which leaves only a pseudo-scalar degree of freedom propagating and removes the other components of the chiral multiplet $\mathcal{A}$. This can be applied to describe goldstone bosons for example \cite{Komargodski:2009rz} or the inflaton \cite{Ferrara:2015tyn, Dalianis:2017okk}. It is important to understand how (\ref{constrintIntro}) can be obtained from  a linearly realised supersymmetry theory in the UV. It was noted in \cite{DallAgata:2016syy} that imposing (\ref{constrintIntro}) is equivalent to three independent constraints:
\beq \overline{X_{NL}}X_{NL} (\mathcal{A}+ \overline{\mathcal{A}}) = 0, \eeq
\beq \overline{X_{NL}}X_{NL}\overline{D_{\dot{\alpha}}\mathcal{A}} = 0, \eeq
\beq \overline{X_{NL}}X_{NL}\overline{D^2 \mathcal{A}} = 0, \eeq
which eliminates the heavy real scalar, the fermion and the auxiliary field separately. These were lifted as three operators added to the Lagrangian with the inconvenience of dealing with higher derivative terms. We shall provide in this work a single  operator that when present in the microscopic theory can give rise to the constraint (\ref{constrintIntro}). This will be based on switching on a D-term to break supersymmetry.

Another issue of interest is the extraction of the Standard Model Higgs $SU(2)$ doublet from a supersymmetric electroweak sector. In its minimal realisation the latter contains two doublet superfields.  We look then for a way to project out the fermionic partners (the higgsinos) and keep only one linear combination of the scalar two Higgs doublets light. We illustrate how this can be achieved easily using two type of operators, one for the $\mu$-like terms and diagonal soft-terms and one for the $B\mu$-term.

As an example of microscopic theory for $D$-term supersymmetry breaking, we consider the original Fayet-Iliopoulos (FI) model \cite{Fayet:1974jb}. In  \cite{Samuel:1982uh}, the parameters region where supersymmetry but not gauge symmetry is broken was considered. It was noted  that replacing the original vector multiplet $V$ in the FI model  by the constrained one, $V \rightarrow V_{NL} $, as described by the above equations leads to the supersymmetry breaking soft masses.  We shall  consider instead the parameter region where both supersymmetry and the gauge symmetry are spontaneously broken leaving in the infrared only the massless goldstino. We shall then show how $V$ flows in the infrared to $V_{NL}  \propto \ov{X_{NL} }X_{NL} $ where $X_{NL}$ is the goldstino nilpotent superfield. For the purpose, we shall illustrate by deriving this result in three different ways: from integrating out heavy modes within the Lagrangian in components fields, identification of the nilpotent superfield in the Ferrara-Zumino supercurrent equation and from integrating out the heavy modes through the superfield equations in the super-unitary gauge.

In section 2, we explain how the supersymmetry algebra fixes the different components of the goldstino multiplet in particular for the nilpotent vector superfield. This result is explicitly derived in section 3 for the case of the FI model in the regime where the only massless degree of freedom is the goldstino. A complete and simple picture is obtained by providing the identification of $X_{NL}$ and $V_{NL}$ by different ways. An important result of this work, the use of a single and simple operator to obtain the minimal constrained superfield, containing a single pseudo-scalar degree of freedom is described in section 4. The discussion about the Higgs sector is in section 5. The conclusions give a summary of the results.

\section{Nilpotent superfield components from supersymmetry algebra}

Integrated out the complex scalar is replaced by a function of the fermionic $\psi$ and the auxiliary field $F$ components of the chiral supermultiplet. 
The supersymmetry transformation reads then:
\bea  \delta_\epsilon\phi (\psi, F) &&= \frac{\partial\phi}{\partial\psi_\alpha} \delta_\epsilon\psi_\alpha +  \frac{\partial\phi}{\partial F} \delta_\epsilon F \nn \\
\epsilon\psi &&= \frac{\partial\phi}{\partial\psi_\alpha} [-i(\sigma^\mu\overline{\epsilon})_\alpha \partial_\mu\phi + \epsilon_\alpha F] - \frac{\partial\phi}{\partial F} (i\overline{\epsilon}\overline{\sigma}^\mu\partial_\mu\psi).\eea
By solving the partial differential equation, we can fix the complex scalar to be:
\beq \phi = \frac{\psi\psi}{2F}.\label{sgoldstino}\eeq
The chiral multiplet can be written as:
\beq X_{NL} = \frac{\psi\psi}{2F} + \sqrt{2}\theta\psi + \theta\theta F,\eeq
which is nilpotent:
\beq X_{NL}^2 = 0.\eeq
Because of the nilpotency constraint, the general form for the Lagrangian without supersymmetric covariant derivatives for this superfield is:
\bea\mathcal{L}_X &&= \int d^4\theta\overline{X_{NL}}X_{NL} + (\int d^2\theta f X_{NL} + h.c.)\nn\\
&&= i\overline{\psi}\overline{\sigma}^\mu\partial_\mu\psi - \partial^\mu (\frac{\overline{\psi\psi}}{2\overline{F}})\partial_\mu(\frac{\psi\psi}{2F}) + \overline{F}F + fF + \ov{f} \overline{F},\label{Eq:L_X}\eea
in which f is a constant. This recovers the Volkov-Akulov action by a field-redefinition.

For a $U(1)$ vector multiplet, fixing a gauge breaks supersymmetry. Thus, after a supersymmetry transformation a new supergauge transformation is required to
go back to the chosen gauge. Choosing the Wess-Zumino gauge, the new supergauge transformation is:
\bea \delta_G V &&= i(\overline{\Lambda} - \Lambda),\nn\\
\Lambda(y) &&= \frac{i}{\sqrt{2}}\theta\sigma^\mu\overline{\epsilon}A_\mu -\theta\theta\frac{i}{\sqrt{2}}\overline{\epsilon\lambda},
\label{gaugetransformationbacktoWZ}
\eea
The combination of the two transformations read then:
\bea \sqrt{2}\delta_{\epsilon + G} A_\mu &&= \epsilon\sigma_\mu\overline{\lambda} + \lambda\sigma_\mu\overline{\epsilon} ,\nn \\
\sqrt{2}\delta_{\epsilon + G}\lambda_\alpha &&= \frac{i}{2} (\sigma^\mu\overline{\sigma}^\nu\epsilon)_\alpha F_{\mu\nu} + \epsilon_\alpha D ,\nn\\
\sqrt{2}\delta_{\epsilon + G} D &&=  -i\overline{\epsilon}\overline{\sigma}^\mu\partial_\mu\lambda + i\partial_\mu\overline{\lambda}\overline{\sigma}^\mu\epsilon .
\eea
Once the gauge group and supersymmetry are broken, we can integrate out $A_\mu$. To do this, we work in the superunitary gauge (i.e. we absorb the Goldstone boson; we shall do this throughout) and, writing the component $A_\mu$ as a function of $\lambda$, $\overline{\lambda}$ and D, the supersymmetric transformation:
\beq \delta_\epsilon A_\mu = \frac{\partial A_\mu}{\partial \lambda_\alpha} [\frac{i}{2\sqrt{2}} (\sigma^\mu\overline{\sigma}^\nu\epsilon)_\alpha F_{\mu\nu} + \frac{1}{\sqrt{2}}\epsilon_\alpha D] + h.c. + \frac{\partial A_\mu}{\partial D} \frac{i}{\sqrt{2}} (-\overline{\epsilon}\overline{\sigma}^\mu\partial_\mu\lambda + \partial_\mu\overline{\lambda}\overline{\sigma}^\mu\epsilon).\eeq
is satisfied if:
\beq A_\mu = \frac{\lambda\sigma_\mu\overline{\lambda}}{D}.\label{heavygaugeboson}\eeq
Note that this is not gauge invariant, as the gauge group is broken; if we restore the would-be Goldstone boson $a$ then we have the relation
$$
A_\mu - \frac{1}{m_{A}} \partial_\mu a= \frac{\lambda\sigma_\mu\overline{\lambda}}{D},
$$
which returns to the above expression when the gauge boson mass $m_A \rightarrow \infty$. 
The corresponding  Lagrangian includes the kinetic term and a Fayet-Iliopoulos term is then:
\bea \mathcal{L}_V &&=  \int d^2\theta \frac{1}{4} W_{NL}W_{NL} + h.c. + \int d^4\theta2\xi V\nn\\
&&=  i\overline{\lambda}\overline{\sigma}^\mu\partial_\mu\lambda -\frac{1}{4}[\partial_\mu(\frac{\lambda\sigma_\nu\overline{\lambda}}{D}) - \partial_\nu(\frac{\lambda\sigma_\mu\overline{\lambda}}{D})][\partial^\mu(\frac{\lambda\sigma^\nu\overline{\lambda}}{D}) - \partial^\nu(\frac{\lambda\sigma^\mu\overline{\lambda}}{D})]+ \frac{1}{2}D^2 + \xi D\nonumber,\\\eea
which is shown to be equivalent to eq. (\ref{Eq:L_X}) if $\xi = \sqrt{2}f$.

\section{Nilpotent goldstino superfield from FI model}

Let us first summarise the Fayet-Ilioupous (FI) model; this allows to fix our notations. It contains two chiral superfields $\Phi_\pm (y, \theta, \thetab) = \phi_\pm (y)+ \sqrt{2} \theta \psi_\pm (y) + \theta \theta F_\pm (y)$, $y^\mu \equiv x^\mu - i \theta \sigma^\mu \thetab$,  with superpotential $W= m \Phi_+ \Phi_-$, a $U(1)$ gauge field and an FI term $\xi $ with the interaction
\begin{align}
\int d^2\theta (\frac{1}{4}W^\alpha W_\alpha + m \Phi_+ \Phi_-) + h.c. + \int d^4\theta \left[  \ov{\Phi_+} e^{2gV} \Phi_+ + \ov{\Phi_-} e^{-2gV} \Phi_- + 2 \xi V \right] \, .
\end{align}
Eliminating the D-term leads to a potential 
\begin{align}
\lagr \supset& -m^2 (|\phi_+|^2 + |\phi_-|^2) - \frac{1}{2}( \xi + g |\phi_+|^2 - g |\phi_-|^2)^2 \, .
\end{align}
We consider the case $\xi g > m^2$ where both the $U(1)$ symmetry and supersymmetry are broken. Writing $\phi_- =\frac{1}{\sqrt{2}} (v + h +ia)$,  we have:
\begin{align}
\frac{g^2 v^2}{2} =& \xi g - m^2  \nn\\
D= - \xi + \frac{gv^2}{2} = - \frac{m^2}{g}  \, ,& \qquad F_+^* = -\frac{mv}{\sqrt{2}}\nn\\
|F_+|^2 + \frac{1}{2} D^2 =& \frac{m^2}{2 g^2} ( m^2 + g^2 v^2)  \ .
\end{align}

 The whole spectrum is: two spinors $\psi_-$ and $\tilde{\psi}$ combined to get a Dirac mass $\sqrt{m^2 + g^2v^2}$, one vector $v_\mu$ and the real scalar $h$ of mass $gv$, one complex scalar field $\phi_+$ of mass $\sqrt{2m^2}$ and one massless Goldstone fermion $\tilde{\lambda}$. The fermionic mass eigenstates  are related to the original fields through the re-definition: 
\begin{align}
(m \psi_+ - gv \lambda) \psi_- \propto& \tilde{\psi} \psi_- \nn\\
\rightarrow \twovec[\tilde{\psi},\tilde{\lambda}] =& \frac{1}{\sqrt{m^2 + g^2v^2}} \twomat[m,-gv][gv,m] \twovec[\psi_+,\lambda]  \nn\\
\twovec[\psi_+,\lambda]  =& \frac{1}{\sqrt{m^2 + g^2v^2}} \twomat[m,gv][-gv,m] \twovec[\tilde{\psi},\tilde{\lambda}].
\label{fermionmasseigenstates}\end{align}
Looking at the supersymmetry transformations we have
\begin{align}
\delta \tilde{\lambda} =& \frac{1}{\sqrt{2}} \epsilon_\alpha \frac{1}{\sqrt{m^2 + g^2v^2}} \big[\sqrt{2}gv F_+ + m D  \big] \nn\\
=& -\frac{m}{\sqrt{2}g} \sqrt{ m^2 + g^2 v^2} \epsilon_\alpha + ... \nn\\
\equiv& \tilde{f} \epsilon_\alpha + ...
\end{align}

\subsection {Integrating out in components}

We should then integrate out all of the fields except $\tilde{\lambda}$ and then relate $\lambda$ to $\tilde{\lambda}$. For the rest, we have the gauge boson EOM, the Higgs, $\phi_+$ and two fermion EOMs:
\begin{align}
A_\mu: 0=& \nabla^2 A_\mu + g^2 v^2 (A_\mu - \frac{1}{gv} \partial_\mu a) + g [ -\psi_- \sigma^\mu \ov{\psi}_- + \frac{(m \tilde{\psi} + gv \tilde{\lambda}) \sigma^\mu (m \bar{\tilde{\psi}} + gv \bar{\tilde{\lambda}})}{m^2 + g^2v^2}] \nn\\
\psi_-: 0=& i \sigma^\mu D_\mu \bar{\psi}_- - \sqrt{m^2 + g^2 v^2} \tilde{\psi} + \frac{g(h-ia)}{\sqrt{m^2 + g^2v^2}} (-gv \tilde{\psi} + m \tilde{\lambda}) \nn\\
\tilde{\psi}: 0=& i \sigma^\mu \partial_\mu \bar{\tilde{\psi}} - \sqrt{m^2 + g^2 v^2} \psi_- + \frac{mg A_\mu}{m^2 + g^2 v^2} \sigma^\mu(m\bar{\tilde{\psi}} + gv \bar{\tilde{\lambda}}) \nn\\
& - \frac{g^2 v(h-ia)}{\sqrt{m^2 + g^2v^2}} \psi_- - \frac{\sqrt{2} g}{m^2 + g^2 v^2}  \phi_+^* [ -2 m gv \tilde{\psi} + (m^2 - g^2 v^2) \tilde{\lambda}] \nn\\
\phi_+^*: 0=& -\partial^2 \phi_+ - 2m^2 \phi_+ + \mathrm{scalar\ terms} - \frac{\sqrt{2} g}{m^2 + g^2 v^2}   (m\tilde{\psi} + gv \tilde{\lambda})(-gv \tilde{\psi} + m\tilde{\lambda}) \nn\\
h: 0=& -\partial^2 h - 2g^2 v^2 h + \mathrm{scalar\ terms} + \bigg[ \frac{g}{\sqrt{m^2 + g^2v^2}} \psi_- (-gv \tilde{\psi} + m \tilde{\lambda}) + h.c. \bigg]
\end{align}
We therefore see that 
\begin{align}
A_\mu \sim& \mathcal{O}(\tilde{\lambda}^2)\nn\\
\phi_+ \sim& \mathcal{O}(\tilde{\lambda}^2)\nn\\
\psi_- \sim& \mathcal{O}(\tilde{\lambda}^3) \nn\\
h \sim& \mathcal{O}(\tilde{\lambda}^4) \nn\\
\end{align}
The imaginary part $a$ of $\phi_-$  which is the would-be gauge boson that can be eliminated in the unitary gauge, can also be shown in other gauges to be  of $\mathcal{O}(\tilde{\lambda}^4)$. We can therefore set $h=a=0$ and expand:
\begin{align}
A_\mu =& -\frac{g}{m^2 + g^2 v^2} \tilde{\lambda} \sigma^\mu \bar{\tilde{\lambda}} + ... \nn\\
\phi_+ =& - \frac{ g^2 v }{\sqrt{2} m(m^2 + g^2 v^2)}   \tilde{\lambda}\tilde{\lambda} + \mathcal{O}(\tilde{\lambda}^4) \nn\\
\psi_- =& - \frac{\sqrt{2} g}{(m^2 + g^2 v^2)^{3/2}}  \phi_+^*  (m^2 - g^2 v^2) \tilde{\lambda} + \frac{mg^2 v A_\mu \sigma^\mu \bar{\tilde{\lambda}}}{(m^2 + g^2v^2)^{3/2}}+  i \sigma^\mu \partial_\mu \bar{\tilde{\psi}}  + ... \nn\\
=& - \frac{g^3 v}{m(m^2 + g^2 v^2)^{3/2}} \ov{\tilde{\lambda}}\ov{\tilde{\lambda}} \tilde{\lambda} + ... \nn\\
\tilde{\psi} =& \frac{1}{\sqrt{m^2 + g^2 v^2}}i \sigma^\mu D_\mu \bar{\psi}_- + ... \nn\\
=& -\frac{g^3 v}{m(m^2 + g^2 v^2)^{2}}  i \sigma^\mu \partial_\mu (\tilde{\lambda}\tilde{\lambda} \ov{ \tilde{\lambda}}) + ...
\end{align}
so we finally find
\begin{align}
\lambda = \frac{gv}{\sqrt{m^2 + g^2 v^2}} \bigg[ \tilde{\lambda} + \frac{g^2 }{(m^2 + g^2 v^2)^{2}}  i \sigma^\mu \partial_\mu [(\tilde{\lambda}\tilde{\lambda}) \ov{ \tilde{\lambda}}] + ... \bigg]
\end{align}
Thus we find that in the low energy limit, the degrees of freedom can be parameterised into one chiral multiplet and one vector multiplet, in an obvious notation:
\bea \Phi_{+} (\phi_+, \psi_+, F_+) &&\overset{IR}{ \longrightarrow}  \frac{gv}{\sqrt{m^2 + g^2v^2}}\Phi_+ (\frac{\tilde{\lambda}\tilde{\lambda}}{2\tilde{f}}, \tilde{\lambda}, \tilde{f}) \label{Eq:Phi1component}\\
V (\lambda, v^\mu, D) &&\overset{IR}{ \longrightarrow}  \frac{m}{\sqrt{m^2 + g^2v^2}}V (\tilde{\lambda}, \frac{\tilde{\lambda}\sigma^\mu\overline{\tilde{\lambda}}}{\sqrt{2}\tilde{f}}, \sqrt{2}\tilde{f} ).\label{Eq:Vcomponent}\eea
This corresponds to Eq. (\ref{sgoldstino}) and (\ref{heavygaugeboson}) and the corresponding Lagrangian can be mapped to the Volkov-Akulov action.

\subsection {Integrating out in superspace}

Let us use the superunitary gauge, in which, the chiral superfield $\Phi_-$ is eaten by the gauge field. Then the Lagrangian becomes:
\begin{align}
 \mathcal{L}_{SU} = \int d^2 \theta ( \frac{1}{4} W^\alpha W_\alpha +\frac{1}{\sqrt{2}} mv\Phi_+ ) + h.c. +\int d^4\theta ( \overline{\Phi_+} e^{2gV} \Phi_+ + \frac{1}{2} v^2e^{-2gV} + 2 \xi V ). \label{Eq:FILagrangeSU}
\end{align}
It is then instructive to consider the two limits $m^2\ll g^2v^2$ and $m^2\gg g^2v^2$ separately. The reason is that we shall integrate out one of the superfields \emph{entirely}, via the equations of motion, while leaving the other light, and this only makes sense if there is a hierarchy of masses. From the component calculation, we observe that in the first limit the Goldstino is dominated by $\psi_+ \supset \Phi_+$, while in the second it is dominated by the gaugino; in superfields unsurprisingly we see that in each limit it is the corresponding superfield that remains in the spectrum.

\subsubsection*{Case $\mathbf{m^2\ll g^2v^2}$:}

In this limit, $\psi_+$ dominates the goldstino. We first consider the equation of motion for $V$:
\begin{align}
0 =& \frac{1}{8} (D^\alpha \ov{D}^2 D_\alpha + h.c. ) V + 2 g \overline{\Phi_+} e^{2gV} \Phi_+ - g v^2e^{-2gV} + 2 \xi .
\end{align}
We then use 
\begin{align}
V \supset& \frac{1}{2} \theta^4 D, \qquad D^\alpha \ov{D}^2 D_\alpha \theta^4 = 16
\end{align}
and write
\begin{align}
V =& \theta^4 \frac{1}{2} (-\xi + \frac{gv^2}{2}) + \hat{V} \equiv \theta^4 \frac{1}{2} \delta + \hat{V} \nn\\
W_\alpha =& \theta_\alpha \delta + \hat{W}_\alpha
\end{align}
which, when combined with
\begin{align}
\int d^4 x\int d^2 \theta  \frac{1}{2} \theta^\alpha \hat{W}_\alpha
=& \int d^4 x \int d^4 \theta \hat{V}
\end{align}
substituted back into the action, gives
\begin{align}
\mathcal{L}_{SU} 
=& \int d^2 \theta ( \frac{1}{4} \hat{W}^\alpha \hat{W}_\alpha +\frac{1}{\sqrt{2}} mv\Phi_+ ) + h.c. +\int d^4\theta ( \overline{\Phi_+} e^{2gV} \Phi_+ + \frac{1}{2} v^2e^{-2gV} + (2 \xi + 2\delta) \hat{V} ) \nn\\
& + \frac{1}{2} \delta^2 + \xi \delta \nn\\
=& \int d^2 \theta ( \frac{1}{4} \hat{W}^\alpha \hat{W}_\alpha +\frac{1}{\sqrt{2}} mv\Phi_+ ) + h.c. +\int d^4\theta ( \overline{\Phi_+} e^{2gV} \Phi_+ + \frac{1}{2} v^2e^{-2gV} + g v^2  V) \nn\\
& + \frac{1}{2} \delta^2 + \xi \delta - \frac{1}{2} \delta gv^2 
\end{align}
This action has no linear term in $V$ once we expand the exponential, which will be what we need. The equations of motion are
\begin{align}
0 =& 2 \delta + \frac{1}{8} (D^\alpha \ov{D}^2 D_\alpha + h.c. ) \hat{V} + 2 g \overline{\Phi_+} e^{2gV} \Phi_+ - g v^2e^{-2gV} + 2 \xi  \nn\\
=& \frac{1}{8} (D^\alpha \ov{D}^2 D_\alpha + h.c. ) \hat{V} + 2 g \overline{\Phi_+} e^{2gV} \Phi_+  + gv^2 (1- e^{-2gV}) \nn\\
\equiv& \Delta + 2 g \overline{\Phi_+} e^{2gV} \Phi_+  + gv^2 (1- e^{-2gV}) 
\end{align}
If we then solve this as a quadratic equation we have
\begin{align}
e^{-2gV} =& \frac{1}{-2gv^2} \bigg[ -gv^2 - \Delta \pm \sqrt{(gv^2 +   \Delta)^2 + 8 g^2 v^2 |\Phi_+|^2} \bigg] \nn\\
=& \frac{(gv^2 +   \Delta)}{2gv^2} \bigg[ 2 + \frac{4g^2 v^2 |\Phi_+|^2}{(gv^2 +   \Delta)^2} + ...\bigg] 
\end{align}
If we neglect the terms with derivatives (i.e. $\Delta$) then we have
\begin{align}
gV =& - \frac{|\Phi_+|^2}{v^2} + 3\frac{|\Phi_+|^4}{v^4} + ...
\end{align}

Let us substitute this back into the action:
\begin{align}
\mathcal{L} =& \int d^2 \theta \frac{1}{\sqrt{2}} mv\Phi_ +  + h.c. + \int d^4 \theta \frac{1}{16} V (D^\alpha \ov{D}^2 D_\alpha + h.c. )V  +  \overline{\Phi_+} e^{2gV} \Phi_+ + \frac{1}{2} v^2e^{-2gV} + 2 \xi V \nn\\
=& \int d^2 \theta \frac{1}{\sqrt{2}} mv\Phi_ + + h.c. + \int d^4 \theta \frac{1}{2} V \bigg[ -2g \overline{\Phi_+} e^{2gV} \Phi_+ + g v^2e^{-2gV} - 2 \xi \bigg] +  \overline{\Phi_+} e^{2gV} \Phi_+ \nn\\
& + \frac{1}{2} v^2e^{-2gV} + 2 \xi V  \nn\\
=& \int d^2 \theta \frac{1}{\sqrt{2}} mv\Phi_ + + h.c. + \int d^4 \theta \, \, \overline{\Phi_+}  \Phi_+ \bigg[ 1 - \frac{m^2}{2g^2 v^2} \bigg] + |\overline{\Phi_+}  \Phi_+|^2 \bigg[ - \frac{1}{v^2} + \frac{3 m^2}{g^2 v^4} \bigg] + ... \label{chiraldominantLagrange}
\end{align}

We note that integrating the gauge field out and retaining the full $\Phi_+$ field only makes sense for $m^2 \ll g^2 v^2$; in this case we have the mass for the $\phi_+$ from the last term  in Eq. (\ref{chiraldominantLagrange}) as
\begin{align}
m_{\phi_+}^2 =& 4 \frac{|F_+|^2}{v^2} = 2 m^2
\end{align}
which is exactly what we found in components. The fact that this equality is only valid for $m^2 \ll g^2 v^2$ (and not true everywhere) is because of the supergauge rotation that we made: we have rotated the $\Phi_+$ and $\Phi_-$ fields among each other. 

Note that (\ref{chiraldominantLagrange}) is of the form of the low energy-limit of the O'Raifeartaigh model and the equations of motion lead to the nilpotency $\Phi_+$ as discussed for instance in \cite{Komargodski:2009rz}.

\subsubsection*{Case $\mathbf{m^2\gg g^2v^2}$:}
In this limit, the gaugino $\lambda$ dominates the goldstino. We can first write the equation of motion for the chiral superfield $\Phi_+$:
\beq 0 = -4\sqrt{2}mv + D^2(e^{2gV}\Phi_+) + \ov{D}^2(\ov{\Phi_+}e^{2gV}).\eeq
This equation is hard to solve. But there is one obvious solution at low energy:
\beq \Phi_+ = c X_{NL}, \qquad V = \ov{X_{NL}}X_{NL}/\Lambda^2,\eeq
in which c  and $\Lambda$ can be determined from the vev of the auxiliary field of $\Phi_+$, $V$ and $X_{NL}$.

\subsection{Nilpotent chiral superfield from Ferrara-Zumino supercurrent}
In the general case, the goldstino is a linear combination of $\psi_+$ and $\lambda$. One easy way to see it is through the Ferrara-Zumino supercurrent. It was noted in \cite{Komargodski:2009rz} that the nilpotent goldstino superfield  controls the non-conservation of the Ferrara-Zumino supercurrent $\mathcal{J}_{\alpha \dot{\alpha}}$  through the equation:
\begin{align}
\ov{D}^ {\dot{\alpha}} \mathcal{J}_{\alpha \dot{\alpha}} = D_\alpha X
\end{align}

It was subsequently shown in \cite{Arnold:2012yi} that in the presence of a FI term, $X$ can be formally obtained in a gauge invariant form as :
\begin{align}
X = 4 W - \frac{1}{3} \ov{D}^2 \left[ K + 2 \xi (V + i \Lambda - i \Lambda^\dagger ) \right].\label{FZX}
\end{align}
In our case, the lagrangian (\ref{Eq:FILagrangeSU}) in the super-unitary gauge gives:
\bea W &&= \frac{1}{\sqrt{2}} mv\Phi_+,\\
K &&= \ov{\Phi_+}e^{2gV}\Phi_+ + \frac{1}{2}v^2e^{-2gV}.\eea
Now we compute the eq. (\ref{FZX}):
\bea X && = 2\sqrt{2}mv\Phi_+ - \frac{1}{3}\ov{D}^2 (\ov\Phi_+ e^{2gV}\Phi_+ + \frac{1}{2}v^2e^{-2gV} + 2\xi V) \nn\\
&&= 2\sqrt{2}mv\Phi_+ -\frac{1}{3}\ov{D}^2 \ov{\Phi_+}\Phi_+ -\frac{1}{3}\ov{D}^2V (-gv^2 + 2\xi) + ...\nn\\
&&= \frac{4\sqrt{2}}{3}mv\Phi_+ - \frac{2}{3}\frac{m^2}{g}\ov{D}^2V + ..., \label{XUV}\eea
in which ... denotes higher order term in the expansion of $e^{\pm 2gV}$. The reason we can neglect them in the IR is that the $\theta$ component contains higher dimension operator than single fermion. We now focus on the $\theta$ component of eq. (\ref{FZX}):
\beq X|_{\theta} = \frac{8}{3}mv\psi_+ + \frac{8}{3}\frac{m^2}{g}\lambda.\label{X|theta}\eeq
Compared to the previous results eq.(\ref{fermionmasseigenstates}), we can identify this as being proportional to $\tilde{\lambda}$. In the IR, the Lagrangian contains only one goldstino with supersymmetry breaking scale:
\beq \tilde{f} = -\frac{m}{\sqrt{2}g}\sqrt{m^2 + g^2v^2}.\eeq
We know that the Lagrangian becomes that of Volkov-Akulov at low energy :
\beq \mathcal{L}_{VA} = \int d^4\theta \ov{X_{NL}}X_{NL} + (\int d^2\theta -\tilde{f}X_{NL} + h.c.),\eeq
in which the nilpotent chiral superfield $X_{NL}$ contains the goldstino:
\beq X_{NL} = \frac{\tilde{\lambda}\tilde{\lambda}}{2\tilde{f}} + \sqrt{2}\theta\tilde{\lambda} + \theta\theta \tilde{f},\eeq
Putting the Volkov-Akulov action into eq. (\ref{FZX}), we can identify:
\beq X = -\frac{8\tilde{f}}{3}X_{NL}.\label{XIR}\eeq
By matching eq. (\ref{XUV}) and (\ref{XIR}) we obtain:
\beq 2\sqrt{2}mv\Phi_+ - \frac{1}{3}\ov{D}^2 (\ov\Phi_+ e^{2gV}\Phi_+ + \frac{1}{2}v^2e^{-2gV} + 2\xi V) \rightarrow  -\frac{8\tilde{f}}{3}X_{NL}.\eeq
The $\theta$ component gives the linear combination of the goldstino:
\beq \tilde{\lambda} = \frac{gv\psi_+ + m\lambda}{\sqrt{m^2 + g^2v^2}}. \eeq
The vev of the auxiliary fields can fix the low energy parameterization of $\Phi_+$ and $V$:
\bea \Phi_+ && \rightarrow \frac{gv}{\sqrt{m^2 + g^2v^2}}X_{NL}\\
V && \rightarrow -\frac{g}{m^2 + g^2v^2}\ov{X_{NL}}X_{NL}
\eea
Both equations of motion and Ferrara-Zumino supercurrent shows that the massive vector multiplet can be parameterized as $\ov{X_{NL}}X_{NL}$ in the infrared. In the following sections, we will use the gauge invariant superfield $W_{NL}^\alpha$ to write the coupling to the visible sector:
\beq W_{NL}^\alpha = \frac{1}{4\sqrt{2}f}\overline{D}^2D^\alpha(\overline{X_{NL}}X_{NL}),\eeq
where $f = |F_{X_{NL}}|$. In components, this reads:

\bea W_{NL}^\alpha = &&\tilde{\lambda}_\alpha + \theta_\alpha\tilde{D} + \frac{i}{2} (\sigma^\mu\overline{\sigma}^\nu\theta)_\alpha \tilde{F}_{\mu\nu} + i\theta\theta (\sigma^\mu\partial_\mu\overline{\tilde{\lambda}})_\alpha,
\eea
where:
\bea 
\tilde{\lambda}_\alpha &&= -\frac{\overline{F_X}}{f}\psi_\alpha - \frac{i}{f}\partial_\mu\phi (\sigma^\mu\overline{\psi})_\alpha\\
\tilde{D} &&= -\frac{\sqrt{2}F_X\overline{F_X}}{f} + \sqrt{2}\partial^\mu\overline{\phi}\partial_\mu\phi - \frac{i}{\sqrt{2}}\overline{\psi}\overline{\sigma^\mu}\partial_\mu\psi - \frac{i}{\sqrt{2}} \psi\sigma^\mu\partial_\mu\overline{\psi}\\
\tilde{A^\mu} &&= -\frac{\psi\sigma^\mu\ov{\psi} + i\ov{\phi}\partial^\mu\phi - i\phi\partial^\mu\ov{\phi}}{\sqrt{2}f}.
\eea

Finally, note that using $X_{NL}^2 = 0$ and ${X_{NL}} D^\alpha X_{NL} = 0$, it is easy to show that $W_{NL}^\alpha$ satisfies
\beq 
\overline{X_{NL}}X_{NL}W_{NL}^\alpha = 0.
\eeq

\section{The minimal constrained superfield}

 We shall describe now the use of the FI goldstino nilpotent superfield introduced above in order to project out all but one degrees of freedom of a chiral superfield $\mathcal{A}^a$. The latter is in the adjoint representation of a gauge symmetry group with field strength superfield $W^a_\alpha$.

We consider the gauge invariant interaction between $\mathcal{A}^a$, $W^a_\alpha$ and $W_{NL}^\alpha$:
\begin{align}
 -\frac{m_D}{f}\int d^2\theta W_{NL}^\alpha W^a_\alpha\mathcal{A}^a 
= -\frac{m_D}{4\sqrt{2}f^2}\int d^2\theta\overline{D}^2D^\alpha(\overline{X_{NL}}X_{NL})W^a_\alpha\mathcal{A}^a.\label{LDG}
\end{align}
Writing the expansion in components, we  find that the fermions $\lambda^a$ from $W^a$ and $\chi^a$ from $\mathcal{A}^a$ combine to make a Dirac fermion of mass $m_D$; the above operator was studied for that purpose in \cite{Goodsell:2014dia}.  Moreover, using the equations of motion to solve the  $D$-term in $W^a_\alpha$, the real part of the scalar in $\mathcal{A}^a$ obtains a mass $2|m_D|$. Therefore  all of these states decouple from the low energy theory in the limit $m_D \rightarrow \infty$. The remaining propagating light degrees of freedom are the goldstino, the gauge boson in $W^a_\alpha$ and the imaginary part of the scalar in $\mathcal{A}^a$. We shall show how this decoupling is described by constraint equations.

First, the equation of motion to the $\mathcal{A}^a$ immediately yields:
\beq 
\overline{D}^2D^{\alpha}(\overline{X_{NL}}X_{NL})W^a_{\alpha} = 0.
\eeq
We can multiply by $\overline{X_{NL}}X_{NL}D_\beta$ to the left hand side, then using the non-zero property of the $D\overline{D}^2D(\overline{X_{NL}}X_{NL})$ and the nilpotency $XD^\alpha X = 0$, we find:
\beq 
\overline{X_{NL}}X_{NL}W_\alpha^a = 0,
\eeq
which projects out the gaugino in $W^a_\alpha$, as expected since it has a large Dirac mass.

Next, we use the equation of motion of $W^a_\alpha$ to find:
\beq 
D_\alpha\overline{D}^2D^\alpha(\overline{X_{NL}}X_{NL}) (\mathcal{A}^a + \overline{\mathcal{A}^a}) - [\overline{D}^2D^\alpha(\overline{X_{NL}}X_{NL})D_\alpha\mathcal{A}^a + h.c.] = 0.
\label{eomW}
\eeq
We can multiply by $\overline{X_{NL}}X_{NL}$ to the left hand side and get rid of the second term using the nilpotency of $X_{NL}$ to obtain the constraint:
\beq
\overline{X_{NL}}X_{NL} (\mathcal{A}^a + \overline{\mathcal{A}^a}) = 0,
\label{yougotit}
\eeq
which eliminates the real part of the scalar.

We can also plug into the l.h.s. of Eq.(\ref{eomW}) $\overline{X_{NL}}X_{NL}D_\beta$ leading to:
\beq\overline{X_{NL}}X_{NL}D_\alpha\mathcal{A}^a = 0,\eeq
which eliminates the fermion $\chi^a$. In a similar way, we can also obtain the constraint that leads to the elimination of the auxiliary field:
\beq\overline{X_{NL}}X_{NL}D^2\mathcal{A}^a = 0,\eeq

For the case of $U(1)$, the equation (\ref{LDG}) can describe an axion superfield coupled to the kinetic mixing between two different U(1) vector multiplets, which makes saxion and axino massive and leaves axion light. More precisely:
\beq \mathcal{L}_{axion} = \frac{1}{f_A} \int d^2\theta W_{NL}^\alpha W_\alpha^{U(1)}A\eeq
in which $A$ is the axion superfield and $W_\alpha^{U(1)}$ is abelian vector superfield. $f_A$ is the decay constant for the axion.

Comparing with (\ref{LDG}) shows that the axion coupling operator is exactly the same as the mass operator for the $U(1)$ Dirac gaugino and the singlet chiral superfield sBino $\Sigma^1$ is identified with the axion superfield. This also leads to a relation between the supersymmetry breaking mediation scale and the axion symmetry breaking scale:
\beq 
m_D \sim \frac{f}{\Lambda} \sim \frac{f}{f_A} \rightarrow \Lambda \sim f_A.
\eeq

The CP-odd scalar $a$ remains massless as expected as an axion, with a coupling:
\beq
\frac{a}{f_A}\epsilon^{\mu\nu\rho\sigma}F^{U(1)}_{\mu\nu}F^{NL}_{\rho\sigma},
\eeq
which shows the corresponding coupling of goldstini to the axion due to a kinetic mixing between the $U(1)$ and a Fayet-Iliopoulos type $U(1)$.

\section{Constrained superfield for Higgs sector}

Given two chiral doublets $H_{1,2}$ carrying opposite $U(1)$ charges, one can write the operator \cite{Nelson:2015cea}:
\bea 
\mathcal{O}_{H _{jj}}&&= \frac{a_{jj} m_{H} }{8\sqrt{2}f}\int d^2\theta\overline{D}^2(D^\alpha V_{NL}D_{\alpha}H_i)H_j\nonumber\\
&&= -\frac{a_{jj} m_H}{2\sqrt{2}f}\int d^2\theta W_{NL}^\alpha (D_{\alpha}H_i) H_j + ...,
\eea
for $i,j=1,2$ where ... represent extra terms that do not contribute to the superpotential (for a different approach, see for example \cite{Antoniadis:2010hs}). Clearly this leads to a Dirac mass $\frac{1}{2}a_{ii} m_H \tilde{H_1} \tilde{H_2} $ for the fermionic modes $\tilde{H_1} \tilde{H_2} $ and to a mass $|a_{ii}|^2 |m_H|^2 |{H_i}|^2$ for the complex scalar in $H_i$ while leaving massless the scalar in $H_j$.  Taking the limit of a large scale supersymmetry breaking leaves at low energy only the scalar component in $H_j$. This limit is described in the constrained superfield language as imposing:
\bea 
\overline{X_{NL}}X_{NL} D_\alpha H_i &&= 0;\nonumber\\
\overline{X_{NL}}X_{NL} D_\alpha H_j &&= 0;\nonumber\\
\overline{X_{NL}}X_{NL} H_j &&= 0\label{1HD}
\eea
which can be obtained using the equations of motion as was done in the previous section.

One interesting application of this operator is to extract the Standard Model Higgs-like doublet from the minimally supersymmetric extended electroweak sector that comes with two Higgs doublets $H_1$ and $H_2$ with opposite gauge charge. The mass of the two Higgsinos and one complex scalar should be heavy, while leaving one light complex higgs (same generalisation to doublet). Both mass eigenstates should be a linear combination of $H_1$ and $H_2$ in order to give the correct Yukawa couplings. For this, we need to supplement it with the additional operator:

\bea 
\mathcal{O}_{H _{12}} &&= -\frac{a^2_{12} m_{H}^2 }{2f^2}\int d^2\theta  W_{NL} W_{NL} H_1 H_2
\label{DoubleW}
\eea
which generates an off-diagonal mass for the scalars $a^2_{12} m_{H}^2 H_1 H_2$. The Higgs mass matrix becomes:
\beq
m_H^2
  \begin{pmatrix}
    a_{11}^2 & a_{12}^2 \\
    a_{12}^2 & a_{22}^2
  \end{pmatrix}\eeq
  One simple way to realize a light eigenstate is to assume $a_{11} = 0$ and $a_{12} \ll a_{22}$. Thus we can take $a_{22}$ to infinity and only retain $\mathcal{O}_{H _{22}}$. The corresponding constraints are exactly Eq. (\ref{1HD}). However, this will cause the problem of large $\tan\beta$. In the more general case, we require $a_{11} a_{22}-a^2_{12} =0$ and derive the equation of motion to either $H_i$:
\beq \frac{1}{4}\ov{D}^2\ov{H_i} = -\frac{a_{ii}m_H}{2\sqrt{2}f}W^\alpha_{NL}D_\alpha H_j - \frac{a_{jj}m_H}{2\sqrt{2}f}D_\alpha(W^\alpha_{NL}H_j) - \frac{a_{12}^2m_H^2}{2f^2}W_{NL}W_{NL}H_j, \label{eomHiggs}\eeq
in which the l.h.s is from the kinetic term. We include it since the F-term of $H_i$ contributes to the mass term of $h_j$. Then we project $X_{NL}\ov{X_{NL}}$, $X_{NL}\ov{X_{NL}}D_\beta$ and $X_{NL}\ov{X_{NL}}D^2$ respectively onto eq. (\ref{eomHiggs}):
\bea \frac{1}{4}X_{NL}\ov{X_{NL}} \ov{D}^2\ov{H_i} &&= \frac{a_{jj}m_H}{2\sqrt{2}f} X_{NL}\ov{X_{NL}}D^\alpha W_{NL\alpha}H_j;\label{projectXX}\\
  0 &&= X_{NL}\ov{X_{NL}}D^\alpha W_{NL\alpha} (a_{ii} + a_{jj}) D_\beta H_j;\label{projectXXD}\\
  0 &&= \frac{a_{ii}m_H}{2\sqrt{2}f}X_{NL}\ov{X_{NL}}D^\alpha W_{NL\alpha}D^2 H_j + \frac{a_{12}^2m_H^2}{2f^2}X_{NL}\ov{X_{NL}}(D^\alpha W_{NL\alpha})^2H_j.\label{projectXXDD}\nn\\
  \eea
  Eq. (\ref{projectXXD}) gives the constraints for the Higgsino:
  \beq \overline{X_{NL}}X_{NL} D_\alpha H_1 = \overline{X_{NL}}X_{NL} D_\alpha H_2 = 0.\eeq
  Applying eq. (\ref{projectXX}) to eq. (\ref{projectXXDD}) gives the constraint for the heavy higgs:
\beq \overline{X_{NL}}X_{NL} (a_{12}^2 H_j + a_{ii}^2\ov{H_i}) = 0.\eeq
If we use the relation $a_{11} a_{22}-a^2_{12} =0$, this is equivalent to
\beq \overline{X_{NL}}X_{NL} (a_{11} H_1 + a_{22}\ov{H_2}) = 0. \eeq

\section {Conclusions}

The goldstino nilpotent superfield is a common tool to write constraints that project out some components of other chiral or vector superfields. Clearly, it is useful to know if there are consistent microscopic origins of each of such constraints. And vice-versa, it is also useful to know which constraints are obtained when taking some decoupling limits of a given theory leading to non-linearly realised supersymmetry. Along this line, we have considered the FI model in a regime where both gauge symmetry and supersymmetry are spontaneously broken, the latter by the combination of both the FI term and an induced F-term. This is a very simple model, with both a pedagogical insight on mechanisms of supersymmetry breaking and possible applications in phenomenology, which  has not been treated in depth in the existing literature. After reviewing the basic knowledge of the model, we proceeded to illustrate in detail how the goldstino  appears to be embedded in a nilpotent superfield. We have worked out the results in different ways leading to a consistent picture that is easy to understand. First, starting from the Lagrangian in components, we have exhibited how the different components of the superfields can be expressed as functions of one goldstone fermion. Then, working directly in superspace, we were able to follow how the nilpotent superfield emerges at low energies. 

As an application which motivated this work, we have first shown how the model allows a minimal constrained superfield which contains only one scalar degree of freedom to be easily obtained. The necessary constraints to eliminate the other degrees of freedom are all embedded in a single operator $\int d^2\theta\overline{D}^2D^\alpha(\overline{X_{NL}}X_{NL})W^a_\alpha\mathcal{A}^a$ involving our goldstino superfield and can be obtained from a microscopic theory in the presence of an effective D-term breaking. We have then discussed how similar operators can play a role in models of axions/axinos and supersymmetric models of electroweak symmetry breaking. Different applications of the resulting minimal constrained superfields can be advocated. It will be interesting to investigate in the future if the the presence of additional sectors in the theory, as those  necessary to write the above mentioned  operator and which contain gauge vector bosons, can play a role in these cases. 

\vskip.1in
\noindent
{\bf Acknowledgments}

\noindent We are grateful to I. Antoniadis, J. P. Derendinger and E. Dudas for useful discussions.  This work  is supported by the Labex ``Institut Lagrange de Paris'' (ANR-11-IDEX-0004-02,  ANR-10-LABX-63) and by the Agence Nationale de Recherche under grant ANR-15-CE31-0002 ``HiggsAutomator''. K.B acknowledges also the support of the the European Research Council (ERC) under the Advanced  Grant Higgs@LHC (ERC-2012-ADG20120216-321133).



\begin{thebibliography} {99}
\bibitem{Rocek:1978nb}
  M.~Ro\v{c}ek,
  ``Linearizing the Volkov-Akulov Model,''
  Phys.\ Rev.\ Lett.\  {\bf 41} (1978) 451.

\bibitem{Volkov:1973ix} 
  D.~V.~Volkov and V.~P.~Akulov,
  ``Is the Neutrino a Goldstone Particle?,''
  Phys.\ Lett.\  {\bf 46B}, 109 (1973).
  
\bibitem{Casalbuoni:1988xh}
  R.~Casalbuoni, S.~De Curtis, D.~Dominici, F.~Feruglio and R.~Gatto,
  ``Nonlinear Realization of Supersymmetry Algebra From Supersymmetric Constraint,''
  Phys.\ Lett.\ B {\bf 220} (1989) 569.


\bibitem{Komargodski:2009rz}
  Z.~Komargodski and N.~Seiberg,
  ``From Linear SUSY to Constrained Superfields,''
  JHEP {\bf 0909} (2009) 066
  [arXiv:0907.2441 [hep-th]].
  
\bibitem{Lindstrom:1979kq}
  U.~Lindstrom and M.~Ro\v{c}ek,
  ``Constrained Local Superfields,''
  Phys.\ Rev.\ D {\bf 19} (1979) 2300.

\bibitem{Ivanov:1978mx}
  E.~A.~Ivanov and A.~A.~Kapustnikov,
  J.\ Phys.\ A {\bf 11} (1978) 2375.
  doi:10.1088/0305-4470/11/12/005
  
\bibitem{Ivanov:1982bpa}
  E.~A.~Ivanov and A.~A.~Kapustnikov,
  J.\ Phys.\ G {\bf 8} (1982) 167.
  doi:10.1088/0305-4616/8/2/004
  
\bibitem{Samuel:1982uh}
  S.~Samuel and J.~Wess,
  ``A Superfield Formulation of the Nonlinear Realization of Supersymmetry and Its Coupling to Supergravity,''
  Nucl.\ Phys.\ B {\bf 221} (1983) 153.


\bibitem{Bandos:2016xyu}
  I.~Bandos, M.~Heller, S.~M.~Kuzenko, L.~Martucci and D.~Sorokin,
  JHEP {\bf 1611} (2016) 109
  doi:10.1007/JHEP11(2016)109
  [arXiv:1608.05908 [hep-th]].
  
\bibitem{Cribiori:2017ngp} 
  N.~Cribiori, G.~Dall'Agata and F.~Farakos,
  ``From Linear to Non-linear SUSY and Back Again,''
  JHEP {\bf 1708}, 117 (2017)
  [arXiv:1704.07387 [hep-th]].

\bibitem{Buchbinder:2017qls}
  E.~I.~Buchbinder, J.~Hutomo, S.~M.~Kuzenko and G.~Tartaglino-Mazzucchelli,
  Phys.\ Rev.\ D {\bf 96} (2017) no.12,  126015
  doi:10.1103/PhysRevD.96.126015
  [arXiv:1710.00554 [hep-th]].
  
\bibitem{GarciadelMoral:2017vnz} 
  M.~P.~Garcia del Moral, S.~Parameswaran, N.~Quiroz and I.~Zavala,
  ``Anti-D3 branes and moduli in non-linear supergravity,''
  JHEP {\bf 1710}, 185 (2017)
  [arXiv:1707.07059 [hep-th]].
  

\bibitem{Luo:2009ib}
  H.~Luo, M.~Luo and S.~Zheng,
  ``Constrained Superfields and Standard Realization of Nonlinear Supersymmetry,''
  JHEP {\bf 1001} (2010) 043
  [arXiv:0910.2110 [hep-th]].
  
\bibitem{Liu:2010sk}
  H.~Liu, H.~Luo, M.~Luo and L.~Wang,
  ``Leading Order Actions of Goldstino Fields,''
  Eur.\ Phys.\ J.\ C {\bf 71} (2011) 1793
  [arXiv:1005.0231 [hep-th]].
  
\bibitem{Kuzenko:2010ef}
  S.~M.~Kuzenko and S.~J.~Tyler,
  ``Relating the Komargodski-Seiberg and Akulov-Volkov actions: Exact nonlinear field redefinition,''
  Phys.\ Lett.\ B {\bf 698} (2011) 319
  [arXiv:1009.3298 [hep-th]].
  
\bibitem{Ferrara:2015tyn}
  S.~Ferrara, R.~Kallosh and J.~Thaler,
  ``Cosmology with orthogonal nilpotent superfields,''
  Phys.\ Rev.\ D {\bf 93} (2016) no.4,  043516
  [arXiv:1512.00545 [hep-th]].
  
\bibitem{Dalianis:2017okk} 
  I.~Dalianis and F.~Farakos,
  ``Constrained superfields from inflation to reheating,''
  Phys.\ Lett.\ B {\bf 773}, 610 (2017)
  [arXiv:1705.06717 [hep-th]].
  
  
\bibitem{DallAgata:2016syy}
  G.~Dall'Agata, E.~Dudas and F.~Farakos,
  ``On the origin of constrained superfields,''
  JHEP {\bf 1605} (2016) 041
  [arXiv:1603.03416 [hep-th]].
  
 
\bibitem{Fayet:1974jb}
  P.~Fayet and J.~Iliopoulos,
  ``Spontaneously Broken Supergauge Symmetries and Goldstone Spinors,''
  Phys.\ Lett.\  {\bf 51B} (1974) 461.
  
\bibitem{Arnold:2012yi} 
  D.~Arnold, J.~P.~Derendinger and J.~Hartong,
  ``On Supercurrent Superfields and Fayet-Iliopoulos Terms in N=1 Gauge Theories,''
  Nucl.\ Phys.\ B {\bf 867}, 370 (2013)
  [arXiv:1208.1648 [hep-th]].
  
\bibitem{Goodsell:2014dia}
  M.~D.~Goodsell and P.~Tziveloglou,
  ``Dirac Gauginos in Low Scale Supersymmetry Breaking,''
  Nucl.\ Phys.\ B {\bf 889} (2014) 650
  [arXiv:1407.5076 [hep-ph]].

\bibitem{Nelson:2015cea}
  A.~E.~Nelson and T.~S.~Roy,
  ``New Supersoft Supersymmetry Breaking Operators and a Solution to the $\mu$ Problem,''
  Phys.\ Rev.\ Lett.\  {\bf 114} (2015) 201802
  [arXiv:1501.03251 [hep-ph]].
 
\bibitem{Antoniadis:2010hs}
  I.~Antoniadis, E.~Dudas, D.~M.~Ghilencea and P.~Tziveloglou,
  ``Non-linear MSSM,''
  Nucl.\ Phys.\ B {\bf 841} (2010) 157
  [arXiv:1006.1662 [hep-ph]].
  
  
   
\end{thebibliography}
\end{document}